\documentclass[twocolumn,prd,superscriptaddress,floatfix,showpacs,nofootinbib]{revtex4}

\usepackage{amsmath}
\usepackage{amssymb}
\usepackage{graphicx}
\usepackage{epsf}
\usepackage{epsfig}
\usepackage{slashed}
\usepackage[english]{babel}
\usepackage[latin1]{inputenc}
\usepackage{rotating}
\usepackage{subfigure}
\usepackage{color}
\usepackage{slashed}
\usepackage{wrapfig}
\usepackage{subdepth}
\usepackage{bbm}
\usepackage{hyperref}
\usepackage{txfonts}

\newcommand{\tr}[1]{\operatorname{tr}{#1}}
\newcommand{\mywidth}{6.0cm}

\graphicspath{ {./Figures/} }

\begin{document}

\title{Early quark production and approach to chemical equilibrium}

\author{D. Gelfand}
 \email[]{d.gelfand@thphys.uni-heidelberg.de}
 \affiliation{Institut f\"{u}r Theoretische Physik, Universit\"{a}t Heidelberg, Philosophenweg 16, 69120 Heidelberg, Germany}
 \affiliation{Institut f\"{u}r Theoretische Physik, Technische Universit\"{a}t Wien, Wiedner Hauptstrasse 8-10/136, 1040~Wien, Austria}

\author{F. Hebenstreit}
 \email[]{hebenstreit@itp.unibe.ch}
 \affiliation{Albert\:Einstein\:Center,\:Institut\:f\"{u}r\:Theoretische\:Physik,\:Universit\"{a}t\:Bern,\:Sidlerstrasse\:5,\:3012\:Bern,\:Switzerland}

\author{J. Berges}
 \email[]{j.berges@thphys.uni-heidelberg.de}
 \affiliation{Institut f\"{u}r Theoretische Physik, Universit\"{a}t Heidelberg, Philosophenweg 16, 69120 Heidelberg, Germany}

\begin{abstract}
We perform real-time lattice simulations of out-of-equilibrium quark production in non-Abelian gauge theory in 3+1-dimensions. Our simulations include the backreaction of quarks onto the dynamical gluon sector, which is particularly relevant for strongly correlated quarks. We observe fast isotropization and universal behavior of quarks and gluons at weak coupling and establish a quantitative connection to previous pure glue results. In order to understand the strongly correlated regime, we perform simulations for a large number of flavors and compare them to those obtained with two light quark flavors. By doing this we are able to provide estimates of the chemical equilibration time.
\end{abstract}


\maketitle

\section{Introduction}
\label{Introduction}

There is considerable progress in understanding the early stages of the non-Abelian plasma's space-time evolution in relativistic heavy-ion collisions. In fact, early times turn out to be most amenable to a systematic theoretical treatment. For instance, large-scale nonequilibrium lattice gauge theory simulations~\cite{Berges:2014bba} recently identified for the first time which thermalization scenario~\cite{Baier:2000sb} is realized in the limit of very high collision energies, where the running gauge coupling is weak. This is possible since in this case the early-time quantum dynamics can be mapped onto a classical-statistical problem, which can be solved on a computer. The findings have been incorporated in state-of-the-art kinetic descriptions to compute the later stages and thermalization of the quark-gluon plasma~\cite{Kurkela:2015qoa}. Remarkably, characteristic aspects of the weak-coupling results might be even carried over to the strong-coupling regime~\cite{Keegan:2015avk}.

While most real-time lattice simulation studies concentrate so far on pure gauge theory, the understanding of the quark dynamics still poses crucial open questions. The inclusion of 
dynamical quarks became recently possible in real time in 3+1 dimensions due to novel lattice techniques.
Since identical fermions cannot occupy the same state, their quantum nature is highly relevant and a consistent quantum theory of quark production in non-Abelian plasmas is envisaged using real-time lattice simulations.
First successful applications to quantum electrodynamics (QED) in 3+1-dimensions led to lattice simulations of the phenomenon of electron-positron pair production for electric field strengths exceeding the Schwinger limit~\cite{Kasper:2014uaa}. Extending these calculations to
quarks in quantum chromodynamics (QCD) is feasible, providing access to a wealth of phenomenologically
relevant processes. 

As an important step in this direction, we present in this work calculations for QCD with $N_c=2$ colors and different
numbers of light quark flavors $N_f$. While we restrict ourselves to non-expanding systems, for the first time fully $3+1$ dimensional simulations with dynamical quarks are performed, extending earlier estimates from boost-invariant 2+1-dimensional simulations~\cite{Gelis:2005pb} or neglecting backreaction~\cite{Tanji:2015ata}. Our simulations include the backreaction of quarks onto the gluon sector, which is particularly relevant for strongly correlated quarks. While our methods are restricted to sufficiently small values of the gauge coupling
$\alpha_s \equiv g^2/(4\pi)$, backreaction effects are controlled by the product $g^2 N_f$ such that even the weak-coupling limit becomes strongly correlated for a large enough number of flavors. In view of applications to heavy-ion collisions, taking $g^2 N_f$ of order one is expected to be a reasonable assumption and this opens the striking possibility to simulate strong-interaction aspects. We present our simulation results
for both perturbatively small and $\mathcal{O}(1)$ values of the backreaction strength $g^2 N_f$.  

The paper is organized as follows. In Sec.~\ref{sec:method} we describe the lattice simulation method and how we extract observables. There we also discuss different sets of initial conditions with large fields or occupancies. Sec.~\ref{subsec:qcd_isotrop} contains the results for quark production at weak coupling for two light quark favors, while Sec.~\ref{subsec:qcd_largeNf} is devoted to quark backreaction and physics at large $N_f$. We end with conclusions in Sec.~\ref{sec:qcd_conclusions}.

\section{Real-time lattice QCD}
\label{sec:method}

We employ classical-statistical lattice gauge theory including fermions \cite{Kasper:2014uaa} to study the early-time dynamics of QCD. In the following, we restrict ourselves to the gauge group SU(2) with generators $t^a$ and adjoint gauge index $a\in\{1,2,3\}$, which obey the algebra $[t^a,t^b]=i\epsilon^{abc}t^c$.
The traceless and Hermitian Pauli matrices 
\begin{equation}
\sigma_1 = \begin{pmatrix} 0 & 1 \\ 1 & 0 \end{pmatrix} \, , \quad \sigma_2 = \begin{pmatrix} 0 & -i \\ i & 0 \end{pmatrix} \, , \quad \sigma_3 = \begin{pmatrix} 1 & 0 \\ 0 & -1 \end{pmatrix} \, 
\end{equation}
form a basis according to
\begin{equation}
t^a = \frac{\sigma_a}{2} \, .
\end{equation}

\subsection{Equations of motion}
\label{subsec:qcd_formulae}

To manifestly preserve gauge-invariance of the lattice theory, we introduce group-valued link variables $U_{\mu,x}$ which are located between lattice sites $x$ and $x+\hat{\mu}$ and point into the direction of $\hat{\mu}$.
They can be parametrized in terms of the algebra-valued gauge field $A_{\mu,x}=t^aA^a_{\mu,x}$ according to
\begin{equation}
 U_{\mu,n} = e^{iga_{\mu}A_{\mu,x}} = e^{iga_{\mu}t^aA^a_{\mu,x}} \ ,
\end{equation}
where no summation over $\mu$ is implied.
Here, $g$ is the gauge coupling and $a_{\mu}$ is the lattice spacing in the temporal ($a_0=a_t$) and spatial ($a_i=a_s$) directions.
The gauge field lattice action in Minkowski space is then constructed from gauge-invariant plaquette variables
\begin{equation}
 U_{\mu\nu,x} = U_{\mu,x}U_{\nu,x+\hat{\mu}}{U^\dagger}_{\hspace{-0.15cm}\mu,x+\hat{\nu}}{U^\dagger}_{\hspace{-0.15cm}\nu,x} \ , 
\end{equation}
with $U^\dagger_{\mu\nu,x}=U_{\nu\mu,x}$, such that
\begin{equation}
 \label{eq:gauge_action}
 S_g = \frac{2}{g^2}\sum_{x}\left(\sum_{j}\frac{a_s}{a_t}\left[2-\tr{U_{0j,x}}\right]-\sum_{j<k}\frac{a_t}{a_s}\left[2-\tr{U_{jk,x}}\right]\right) \ .
\end{equation}
The plaquette variables $U_{\mu\nu,x}$ encode the non-Abelian field strength tensor $\mathcal{F}_{\mu\nu,x}=\mathcal{F}^a_{\mu\nu,x}t^a$.
In fact, to leading order in the lattice spacing we find
\begin{equation}
 \mathcal{F}^a_{\mu\nu,x} = -\frac{2i}{g a_\mu a_\nu}\tr{[ t^a U_{\mu\nu,x}]} \ .
\end{equation}
The chromoelectric and chromomagnetic field components are given by
\begin{subequations}
\label{elfield_bfield}
\begin{eqnarray}
 E^a_{i,x} &=& -\frac{2i}{g a_t a_s}\tr{[ t^a U_{0i,x}]}\ , \\
 B^a_{i,x} &=& \frac{i}{g a^2_s}\epsilon_{ijk}\tr{[ t^a U_{jk,x}]} \ .
\end{eqnarray} 
\end{subequations}
In the fermionic sector, we employ a gauge-invariant central derivative discretization of the Dirac action
\begin{equation}
 \label{eq:dirac1_action}
 S^0_{\psi}=a_ta^3_s\sum_{x,\mu}\bar{\psi}_x\left( i\gamma^\mu\frac{U_{\mu,x}\psi_{x+\hat{\mu}}-{U^{\dagger}}_{\hspace{-0.15cm}\mu,x-\hat{\mu}}\psi_{x-\hat{\mu}}}{2a_\mu}-m\psi_x \right) \, ,
\end{equation}
along with a pseudoscalar Wilson term $S^W_{\psi}[U,\bar{\psi},\psi]$ in order to resolve the fermion doubling problem \cite{Nielsen:1981hk}
\begin{equation}
 \label{eq:dirac2_action}
 S^W_{\psi}=a_ta^3_s\sum_{x,j}\bar{\psi}_x\left( i\gamma_5\frac{U_{j,x}\psi_{x+\hat{j}}-2\psi_x+{U^{\dagger}}_{\hspace{-0.15cm}j,x-\hat{j}}\psi_{x-\hat{j}}}{2a_s} \right) \, .
\end{equation}
In comparison to a scalar Wilson term as employed in, e.~g.~\cite{Kasper:2014uaa,Aarts:1998td,Borsanyi:2008eu,Saffin:2011kc,Hebenstreit:2013qxa,Buividovich:2015jfa}, the pseudoscalar Wilson term reduces lattice spacing artifacts as discussed in \cite{Berges:2013oba}.
The total action of the SU(2) gauge theory under consideration is then the sum of the three contributions \eqref{eq:gauge_action}, \eqref{eq:dirac1_action} and \eqref{eq:dirac2_action}.
We note that the action is by construction invariant under local gauge transformations
\begin{subequations}
\begin{align}
 U^{\prime}_{\mu,x} &= G_xU_{\mu,x}{G^{\dagger}}_{\hspace{-0.15cm}x+\hat{\mu}} \, , \\
 \psi^{\prime}_x &= G_x\psi_x \, ,
\end{align}
\end{subequations}
with the gauge transformation matrix $G_x\in \text{SU(2)}$.
Taking advantage of the gauge freedom we may employ the temporal-axial gauge condition $U_{0,x}=\mathbf{1}$ to simplify the simulations afterwards.

The equation of motion governing the time evolution of the fermionic degrees of freedom is then given by
\begin{align}
 \label{eq:disct_dirac}
 \psi_{x+\hat{t}} \,&= \psi_{x-\hat{t}}-2ia_t\left(m-\frac{3i}{a_s}\gamma_5\right)\gamma^0\psi_x-\frac{a_t}{a_s}\gamma^0 \nonumber \\
  &\times\sum_j\left[(\gamma^j+\gamma_5)U_{j,x}\psi_{x+\hat{j}}-(\gamma^j-\gamma_5){U^\dagger}_{\hspace{-0.15cm}j,x-\hat{j}}\psi_{x-\hat{j}} \right] \, .
\end{align}
Here, $\psi_x$ can be taken as stochastic spinor fields in the framework of the male/female method \cite{Borsanyi:2008eu} employed below or, equivalently, as mode functions in an expansion of the Dirac field operator \cite{Aarts:1998td}.
On the other hand, the equation of motion for the chromoelectric field is given by
\begin{align}
 \label{eq:eom_gauge}
 E^{a}_{j,x}\,&=E^{a}_{j,x-\hat{t}}+ga_t\operatorname{Re}\tr{\big[F_{x+\hat{j},x}(\gamma^j+\gamma_5)t^aU_{j,x}\big]}+\frac{2ia_t}{ga^3_s} \nonumber \\
 &\times\sum_{i\neq j}\left(\tr{\big[t^a U_{ji,x}\big]}+\tr{\big[t^a U_{j,x}{U^\dagger}_{\hspace{-0.15cm}i,x+\hat{j}-\hat{i}}{U^\dagger}_{\hspace{-0.15cm}j,x-\hat{i}}U_{i,x-\hat{i}}\big]}\right) \ . 
\end{align}
The backreaction of the fermions onto the gauge fields is determined by the statistical propagator
\begin{align}
 \label{eq:stat_prop}
 F_{x,y}=\frac{1}{2}\langle[\psi_x,\bar{\psi}_y]\rangle \, .
\end{align}
The corresponding trace in \eqref{eq:eom_gauge} is taken over Dirac indices and fundamental gauge indices.
The system of dynamic equations is closed by the time evolution equation of the spatial link variables $U_{j,x}$, which is obtained by reversing \eqref{elfield_bfield} and constructing the temporal plaquette from $E^{a}_{j,x}$ according to
\begin{equation}
 U_{0j,x}=\mathbf{1}\sqrt{1-\left(\frac{ga_ta_s}{2}\right)^2\sum_{a}\left(E^{a}_{j,x}\right)^2}+iga_ta_st^{a}E^{a}_{j,x} \ .
\end{equation}
Taking into account the definition of the temporal plaquette $U_{0j,x}$ in temporal-axial gauge, we obtain the evolution equation for the spatial link
\begin{equation}
 U_{j,x+\hat{t}}= U_{0j,x}U_{j,x} \, .
\end{equation}
In addition to the dynamical equations of motion, we have to impose the Gauss constraint in order to simulate in the physical subspace of the theory
\begin{align}
 \label{eq:gauss_law}
 &\sum_{j}\left(E^{a}_{j,x}+\frac{2i}{ga_ta_s}\tr{\big[t^{a}{U^\dagger}_{\hspace{-0.15cm}j,x+\hat{j}}U_{0j,x-\hat{j}}U_{j,x-\hat{j}}\big]}\right) = \nonumber \\
  &\qquad\qquad\qquad\qquad\qquad\qquad -ga_s\operatorname{Re}\tr{\big[F_{x+\hat{t},x}\gamma^0t^a\big]} \ . 
\end{align}
We emphasize that the time evolution conserves the Gauss constraint.
In practice, we enforce the Gauss constraint at initial times via an iterative method and monitor its possible violation due to rounding errors during runtime.

\subsection{Male/female low-cost fermions}
\label{sec:male_female}

The fermionic contribution to the classical-statistical dynamics can be evolved in time via a mode function expansion without further approximations \cite{Aarts:1998td}.
This approach has been successfully applied for low-dimensional systems \cite{Baacke:1998di,Giudice:1999fb,Hebenstreit:2013baa,Hebenstreit:2014rha}, however, its application for three dimensional systems becomes rather expensive as the computational cost scales like the spatial volume squared.
Low-cost fermions may provide a numerically more efficient method by replacing the mode functions by an ensemble of fields of different ``gender'', denoted as $\psi^{M}_{x}$ (male) and $\psi^{F}_{x}$ (female) \cite{Borsanyi:2008eu}.
In this approach the statistical propagator \eqref{eq:stat_prop}, which governs the backreaction of the fermions onto the gauge fields, is described by 
\begin{equation}
 F_{x,y}\stackrel{!}{=}F^{\text{sto}}_{x,y}=\langle\psi^{M}_{x}\bar{\psi}^{F}_{y}\rangle_{\text{sto}}=\langle\psi^{F}_{x}\bar{\psi}^{M}_{y}\rangle_{\text{sto}} \ , 
\end{equation}
where $\langle\cdots\rangle_{\text{sto}}$ is understood as a stochastic ensemble average over all pairs of male and female spinor fields. 
Convergence to the exact correlator may be achieved, provided that the $\psi^{g}_x$ satisfy the Dirac equation \eqref{eq:disct_dirac} and using that the initial value of the stochastically sampled propagator reproduces the initial conditions for the exact propagator.

To generate initial conditions corresponding to a fermion vacuum at initial times $t_0$, we choose
\begin{equation}
 \psi^{g}_{x=(t_0,\mathbf{x})}=\int{\frac{d^3p}{(2\pi)^3}e^{i{\bf p}{\bf x}}\frac{1}{\sqrt{2}}\sum_{s,i}[u_{s,i,\mathbf{p}}\xi_{s,i,\mathbf{p}}\pm v_{s,i,-\mathbf{p}}\eta_{s,i,\mathbf{p}}]} \ ,
\end{equation}
with free particle spinors $u_{s,i,\mathbf{p}}$ and antiparticle spinors $v_{s,i,\mathbf{p}}$.
Here, $s$ denotes the spin index and $i$ the fundamental gauge index. 
The complex random variables $\xi_{s,i,\mathbf{p}}$ and $\eta_{s,i,\mathbf{p}}$ are sampled according to
\begin{equation}
 \langle\xi_{s,i,\mathbf{p}}{\xi^*}_{\hspace{-0.15cm}s',j,\mathbf{q}}\rangle_{\text{sto}}=\langle\eta_{s,i,\mathbf{p}}{\eta^*}_{\hspace{-0.15cm}s',j,\mathbf{q}}\rangle_{\text{sto}}=(2\pi)^3\delta_{ss'}\delta_{ij}\delta({\bf p}-{\bf q}) \ ,
\end{equation}
whereas all other correlators vanish.
We note that the spinors of different gender differ only by the sign in front of the antiparticle component.
The subsequent time evolution of the fields $\psi^{g}_{x}$ proceeds independently for each member of the ensemble.
In practice, the stochastic average $\langle\cdots\rangle_{\text{sto}}$, which is required for calculating the backreaction onto the gauge fields \eqref{eq:eom_gauge} or for computing fermionic observables, is given by an average over a sufficiently large number $N_{\text{sto}}$ of pairs of male and female spinor fields.

We emphasize again that the computational cost of the mode functions approach scales with the volume of the phase space, thus being proportional to $N^{2d}$ in $d$ dimensions and $N$ being the number of lattice points in each spatial direction. 
Fortunately, the resource requirements of the low-cost approach only scales with the spatial volume times the number of stochastic spinor pairs $N_\text{sto}N^d\ll N^{2d}$, enabling large-scale numerical simulations in three dimensions.

\subsection{Gauge-field initial conditions}
\label{sec:qcd_init}
In the framework of classical-statistical field theory, bosonic correlation functions $\langle O[A]\rangle$ are calculated as ensemble averages by numerically solving the classical field equations and sampling over the initial conditions \cite{Khlebnikov:1996mc,Aarts:2001yn,Polkovnikov:2003,Berges:2007ym,Epelbaum:2014yja}
\begin{equation}
 \langle O[A]\rangle=\int{\mathcal{D}A}\int{\mathcal{D}A_0\mathcal{D}E_0}\,\rho[A_0,E_0]O[A]\,\delta[D_\mu F^{\mu\nu}-j^\nu] \ .
\end{equation}
The initial conditions of the gauge field $A_0$ and the chromoelectric field $E_0$ are sampled according to the Wigner transform of the initial density matrix $\rho[A_0,E_0]$.
On the other hand, the delta function $\delta[D_\mu F^{\mu\nu}-j^\nu]$ enforces the gauge field to obey the Yang-Mills equations \eqref{eq:eom_gauge} and \eqref{eq:gauss_law}.
For further details on the derivation of this equations we refer to \cite{Kasper:2014uaa}.
In the following, we study fermion production for different gluonic initial conditions corresponding to a saturated state of overpopulated gluons or anisotropic classical fields.
In order to fulfill the requirements of the classical-statistical field theory approximation to the underlying quantum dynamics, we have to work at weak gauge coupling and large field amplitudes or occupation numbers~\cite{Khlebnikov:1996mc,Aarts:2001yn,Polkovnikov:2003,Berges:2007ym,Epelbaum:2014yja}.

A corresponding Gaussian density matrix is determined by the chromomagnetic and chromoelectric one-point functions, $\langle B^a_{i,x=(t_0,\mathbf{x})}\rangle$ and $\langle E^a_{i,x=(t_0,\mathbf{x})}\rangle$, along with their two-point correlation functions. The chromomagnetic fields are fully determined by the spatial gauge fields $A^a_{i,x}$ via $B^a_{i,x}=\epsilon^{ijk}F_{jk,x}^a$.
We initialize the gauge field fluctuations to represent a gas of particles with only spatially transverse degrees of freedom, in close similarity to a non-interacting photon gas in quantum electrodynamics with additional internal group indices. At initial time $t_0$, we represent the gauge fields and the chromoelectric fields in momentum space according to
\begin{subequations}
\label{eq:init_fields}
\begin{align}
 A^a_{j,(t_0,\mathbf{q})}&=\sqrt{\frac{f^a_g(|{\bf q}|,t_0)+1/2}{|{\bf q}|}} \sum_\lambda \left[ b^a_{\lambda,{\bf q}}\epsilon_{\lambda,j,{\bf q}} + b^{a,*}_{\lambda,-{\bf q}}\epsilon^*_{\lambda,j,-{\bf q}} \right] \, , \\
 E^a_{j,(t_0,\mathbf{q})}&=i\sqrt{|{\bf q}|\left(f^a_g(|{\bf q}|,t_0)+1/2\right)} \sum_\lambda \left[ b^a_{\lambda,{\bf q}}\epsilon_{\lambda,j,{\bf q}} - b^{a,*}_{\lambda,-{\bf q}}\epsilon^*_{\lambda,j,-{\bf q}}  \right] \, .  
\end{align}
\end{subequations}
Here, $\lambda\in\{1,2\}$ is the polarization index, $\epsilon_{\lambda,j,{\bf q}}$ are the components of normalized polarization vectors orthogonal to the momentum of propagation $\mathbf{q}$, and $f^a_g(|\mathbf{q}|,t_0)$ is the initial gluonic occupation number. 
The polarization vectors for a given momentum $\mathbf{q}$ are constructed numerically according to $\epsilon_{1,\mathbf{q}}=\mathbf{r}\times\mathbf{q}/|\mathbf{r}\times\mathbf{q}|$ and $\epsilon_{2,\mathbf{q}}=\epsilon_1\times\mathbf{q}/|\mathbf{q}|$, with a random vector $\mathbf{r}$.
The complex random numbers $b^a_{\lambda,{\bf q}}$ are chosen such that the only non-vanishing, connected two-point correlation functions read
\begin{subequations}
\begin{align}
 \langle A^a_{i,(t_0,\mathbf{p})}A^b_{j,(t_0,\mathbf{q})}\rangle_{\mathcal{C}}&=\frac{1}{|\mathbf{p}|}\left(f^a_g(|{\bf p}|,t_0)+1/2\right)\mathcal{P}_{ij}\,\delta({\bf p}+{\bf q})\,\delta^{ab} \, , \\ 
 \langle E^a_{i,(t_0,\mathbf{p})}E^b_{j,(t_0,\mathbf{q})}\rangle_{\mathcal{C}}&=|\mathbf{p}|\left(f^a_g(|{\bf p}|,t_0)+1/2\right)\mathcal{P}_{ij}\,\delta({\bf p}+{\bf q})\,\delta^{ab} \, , 
\end{align}
\end{subequations}
with the transverse projector $\mathcal{P}_{ij}=\delta_{ij}-p_ip_j/|\mathbf{p}|^2$.
We emphasize that disconnected contributions of the two-point correlation functions, which are initialized as macroscopic classical fields in some of our scenarios, have been omitted here for notational simplicity.
In the following, we further discuss our initial conditions corresponding to a state of overpopulated gluons or anisotropic classical fields undergoing a rapid decay due to instabilities and particle production. 

\subsubsection*{Overpopulated gluons}

The overpopulation scenario is realized by an initial distribution of the gluonic occupation numbers
\begin{equation}
f_g(|{\bf p}|,t_0) = \frac{1}{g^2}\Theta(|{\bf p}|-Q_s) \, ,
\end{equation}
where the Heaviside function ensures that gluons populate all infrared modes up to the characteristic scale $Q_s$ with a parametrically large occupancy of the order of $1/g^2$.
We emphasize that this initial distribution is isotropic in momentum space and in all color indices.
In what follows, we refer to this as ``fluctuation'' initial condition.
 
The approach to thermal equilibrium from this kind of initial conditions is marked by the transport of energy and particles to short length scales and by an overall reduction in the total number of gluons. 
The latter may also be seen by a parametric estimate, which is valid in the weak coupling limit $g^2\ll1$.
In fact, integration of the initial state distribution $f_g(|{\bf p}|,t_0)$ yields $\varepsilon\sim Q_s^4/g^2$ for the total quasi-particle energy and $n_g^{(0)}\sim Q_s^3/g^2$ for the total quasi-particle number. 
Because of energy conservation, the final temperature of the thermal gluon gas is supposed to be $T \sim Q_s/\sqrt{g}$, indicating that the number of particles in the thermal ensemble should scale as $n_g^{\text{th}} \sim T^3 \sim Q_s^3/g^{3/2}$.
Accordingly, we find at weak coupling $n_g^{(0)}>n^{\text{th}}_g$, elucidating the notion of an overpopulated initial state at weak coupling.

\subsubsection*{Plasma instability}

The evolution of the overpopulated initial state will be compared to a system where the overpopulation is dynamically generated via a Nielsen-Olesen-type magnetic plasma instability \cite{Nielsen:1978rm,Berges:2011sb}. 
This instability is triggered by an initial chromomagnetic field along the $3$-direction, $\langle B_{i,(t_0,\mathbf{x})}^a \rangle=\delta^{1a}\delta_{i3}B$, where $B$ is the chromomagnetic field strength.
In order to realize this configuration, we initialize our simulations with macroscopic gauge fields \cite{Berges:2011sb}
\begin{equation}
 \label{Bfield}
 \langle A^{a=2}_{1,(t_0,\mathbf{x})}\rangle = \langle A^{a=3}_{2,(t_0,\mathbf{x})}\rangle = \sqrt{\frac{B}{g}} \, .
\end{equation}
In the following, we denote this type of macroscopic gauge fields as Nielsen-Olesen-type or ``condensate'' initial condition.

The initial gauge fields cause the longitudinal field $B(t)$ to perform damped oscillations in time, where the damping is driven by interactions with exponentially growing gluon fluctuations. 
The momentum dependent growth rate of these fluctuations in the linear regime is $\gamma_p = (g\bar{B}-p^2_z)^{1/2}$, with $\bar{B}$ being the time-averaged absolute value of the chromomagnetic field.
In fact, the Nielsen-Olesen-type instability is accompanied by the phenomenon of parametric resonance for this initial condition \cite{Berges:2011sb}. While the latter phenomenon is included in the simulation results, we skip its discussion here since it does not dominate the total particle production and refer to \cite{Berges:2013oba,Berges:2004yj,Berges:2015kfa} for further details. 

\subsubsection*{Flux tube}

According to the ``color glass condensate'' picture of the early phase of heavy-ion collisions \cite{Gelis:2010nm},
the initial coherent color fields form so-called flux tubes, which are regions in space in which both chromoelectric and chromomagnetic fields are aligned in the longitudinal direction.
In fact, a single flux-tube resembles the Nielsen-Olesen-type initial condition plus an additional chromoelectric field $\langle E_{i,(t_0,\mathbf{x})}^a \rangle=\delta^{1a}\delta_{i3}E$, where $E$ is the chromoelectric field strength. The macroscopic chromoelectric field induces damped plasma oscillations and simultaneously creates gluons and quarks via the Schwinger mechanism \cite{Schwinger:1951nm,Casher:1978wy}.
In the following section, we show how the longitudinal flux tubes dissipate into fluctuations and discuss the resulting spectrum of quarks and gluons.

\subsection{Particle numbers}
\label{sec:qcd_partnumb}

The notion of a particle number is uniquely defined only in non-interacting relativistic field theory.
In the following, we define an adiabatic quasi-particle number which coincides with the definition of free particles in the non-interacting limit. 

As our initial conditions \eqref{eq:init_fields} are transverse in momentum space, we choose to enforce this property also in our definition of the occupation number.
Due to the fact that the temporal-axial gauge condition $U_{0,n}=\mathbf{1}$ is incomplete in the sense that it leaves a residual gauge invariance under time-independent gauge transformations \cite{Leibbrandt:1987qv}, we may project onto the transverse degrees of freedom at any instant of time 
\begin{equation}
 \label{eq:coulomb_like}
 \nabla_{\bf x} \cdot {\bf A}^a_{x} = 0 \, .
\end{equation}
We emphasize that this condition will only be fulfilled locally in time but not globally in the sense of a Coulomb-gauge condition. Upon imposing \eqref{eq:coulomb_like}, we are able to project on the two transverse polarization modes per gluon at any instant of time.

In order to read out the occupation number distribution, we transform the gluonic and fermionic variables by a gauge transformation $G_{\perp,x}\in\text{SU(2)}$ such that \eqref{eq:coulomb_like} is fulfilled.
Numerically, this is done via a stochastic overrelaxation algorithm as described in \cite{Cucchieri:2003fb}. 
The transformation matrices $G_{\perp,x}$ computed by this algorithm are then used to transform the variables according to
\begin{subequations}
\begin{align}
 U^{\perp}_{i,x} &= G_{\perp,x}U_{i,x}{G^{\dagger}}_{\hspace{-0.15cm}\perp,x+\hat{i}} \ , \\
 U^{\perp}_{0i,x} &= G_{\perp,x}U_{0i,x}{G^{\dagger}}_{\hspace{-0.15cm}\perp,x} \ , \\
 \psi^{\perp}_x &= G_{\perp,x}\psi_x \, .
\end{align}
\end{subequations} 

Our definition of the fermion occupation number distribution $f_\psi(\mathbf{p},t_0)$ is essentially the same as the one employed in \cite{Berges:2010zv,Berges:2013oba}.
To this end, we use the definition of the momentum space gauge-transformed equal-time statistical propagator
\begin{equation}
 F^{\perp}_{(t,\mathbf{x}),(t,\mathbf{y})}=\frac{1}{2}\langle[\psi^\perp_{(t,\mathbf{x})},\bar{\psi}^\perp_{(t,\mathbf{y})}]\rangle=\int{\frac{d^3p}{(2\pi)^3}e^{i\mathbf{p}(\mathbf{x}-\mathbf{y})}F^\perp(\mathbf{p},t)} \, ,
\end{equation}
and project onto its scalar, pseudoscalar and vector components in Dirac space
\begin{subequations}
\label{eq:dirac_components}
\begin{align}
 F_S(\mathbf{p},t)&=\frac{1}{4}\tr{\left[F^\perp(\mathbf{p},t)\right]} \, , \\
 F_{PS}(\mathbf{p},t)&=\frac{1}{4}\tr{\left[F^\perp(\mathbf{p},t)\gamma^5\right]} \, , \\
 F_V^i(\mathbf{p},t)&=\frac{1}{4}\tr{\left[F^\perp(\mathbf{p},t)\gamma^i\right]} \, ,
\end{align}
\end{subequations}
where the trace is over Dirac, color and flavor indices.
Taking the ratio of the total energy density, which can be expressed in terms of \eqref{eq:dirac_components}, with the single-particle energy serves us as a definition of the fermion occupation number
\begin{equation}
\label{eq:npsilat}
 f_{\psi}({\bf p},t)=\frac{1}{2}-\frac{F_{S}({\bf p},t)\, m+F_{V}^{i}({\bf p},t)\, \bar{p}_{i}+i F_{PS}({\bf p},t)\, {\bf p}^2_{\rm lat} a_s/2}{\sqrt{m^{2}+p_ip^i+\big({\bf p}_{\rm lat}^{2} a_{s}/2\big)^2}} .
\end{equation}
Here, $\bar{p}_i$ is the lattice momentum corresponding to a first-order spatial derivative and ${\bf p}^2_{\rm lat}$ is the usual lattice momentum squared
\begin{subequations}
\begin{align}
 \bar{p}_i&=\frac{1}{a_s}\sin\left(p_i a_s\right) \ , \\
 \mathbf{p}_{\rm{lat}}^2&=\frac{4}{a_{s}^{2}}\sum_{i=1}^{3}\sin^{2}\left(\frac{p_{i}a_{s}}{2}\right) \ , 
\end{align}
\end{subequations}
with $p_i=2\pi n_i/(a_sN)$ and $n_i\in\{0,\cdots,N-1\}$. We define the gluonic occupation number according to
\begin{equation}
 f_g(|{\bf p}|,t) = \frac{\sqrt{\tr{\left\langle A^{b,\perp}_{i,(t,\mathbf{p})}A^{c,\perp}_{j,(t,-\mathbf{p})} \right\rangle} \tr{\left\langle E^{b,\perp}_{i,(t,\mathbf{p})}E^{c,\perp}_{j,(t,-\mathbf{p})} \right\rangle}}}{6}-\frac{1}{2} \, .
\end{equation}
Here, the trace is over both color and polarization indices such that this particle number corresponds to an average over all internal degrees of freedom. 

\section{Isotropization and quark production at weak coupling}
\label{subsec:qcd_isotrop}
We first consider two-color QCD with $N_f=2$ degenerate light quark flavors at weak coupling $g^2=10^{-2}$ on a $64^3$ spatial lattice. Parametrically, the backreaction of quarks onto the gluons is expected to be important either at strong couplings (which would be beyond the range of validity of classical-statistical simulations) or for a larger number of flavors $N_f\gg1$. Even though the backreaction effects are supposed to be small for the parameters chosen we still include it for the reason of consistency, most notably to obey the conservation of the total energy during the simulation.

In the previous section, we introduced three generic types of initial conditions which are potentially relevant for the early-time dynamics in heavy-ion collisions.
In fact, the initial condition corresponding to gluonic overpopulation is isotropic from the very beginning, whereas the initial conditions corresponding to the flux tube or the Nielsen-Olesen-type instability are highly anisotropic.
However, both plasma instabilities and the Schwinger mechanism tend to isotropize an initially anisotropic system at later times. This raises the question of whether the system will become insensitive to details of the initial conditions, or may even exhibit universal properties during its time evolution. Accordingly, we first study the  evolutions starting with the flux tube or Nielsen-Olesen-type initial conditions and focus on quantities which are a measure of the system's anisotropy.

\subsection{Coherent field decay}

\begin{figure}[t]
\begin{center}
\epsfig{file=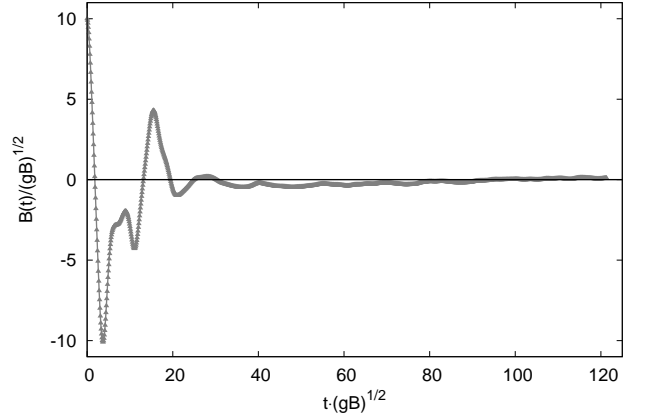, width=\mywidth, angle=270}
\caption{Time evolution of the chromomagnetic field $B(t)$ during the Nielsen-Olesen-type instability for condensate initial conditions.}
\label{fig:bfield_intime}
\end{center}
\end{figure}

We first consider the time evolution of the chromomagnetic field along the $3$-direction
\begin{equation}
 B(t)=\left\langle B^{a=1}_{3}\right\rangle(t)=\frac{1}{2}\sqrt{gB}\left( \left\langle A^{a=2}_1\right\rangle(t) +  \left\langle A^{a=3}_2\right\rangle(t) \right) \, .
\end{equation}
Here $B$ denotes a parameter determining the initial energy density, whereas $B(t)$ evolves as a function of time
with, initially, $B(t_0)=B$ according to \eqref{Bfield}.
In Fig.~\ref{fig:bfield_intime}, we show the time evolution of the chromomagnetic field $B(t)$ for Nielsen-Olesen-type initial conditions.
Starting from a large amplitude, the field exhibits damped oscillations and finally approaches zero at late times.
Due to energy conservation, all the energy initially contained in the one-point function is then transferred to higher correlation functions.

\begin{figure}[t]
\begin{center}
\epsfig{file=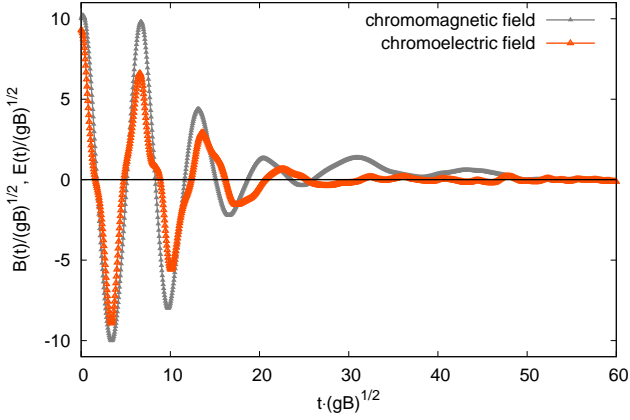, width=\mywidth, angle=270}
\caption{Time evolution of the chromoelectric field $E(t)$ and the chromomagnetic field $B(t)$ for flux tube initial conditions.}
\label{fig:flux_tube_decay}
\end{center}
\end{figure}

A similar behavior is found upon adding a macroscopic initial chromoelectric field to the chromomagnetic field, corresponding to flux tube initial conditions.
In Fig.~\ref{fig:flux_tube_decay} we present the time evolution of the $3$-component of the chromoelectric field $E(t)=\langle E^{a=1}_3\rangle(t)$ along with the corresponding chromomagnetic field $B(t)$.
Again, both macroscopic fields undergo damped oscillations and decay by non-linear interactions and particle production: 
The magnetic field destabilizes gauge fluctuations and causes exponentially growing infrared occupancies of modes \cite{Berges:2011sb,Fujii:2009kb}.
The electric field, on the other hand, directly produces quarks and gluons \cite{Tanji:2011di} and subsequently accelerates them via the non-Abelian Lorentz force.

We observe for both anisotropic scenarios (Nielsen-Olesen-type, flux tube) a depletion of the initially macroscopic fields. In fact, the chromoelectric field vanishes after about $t\simeq 30/\sqrt{gB}$ whereas the decay of the chromomagnetic field takes slightly longer, $t \simeq 50/\sqrt{gB}$.
This suggests that the energy transfer from the chromoelectric sector is more efficient than from the chromomagnetic one. This observation may be interpreted in analogy to the Lorentz force in electrodynamics:
Electric fields perform work, resulting in an increase of the particle's momentum, whereas magnetic fields alter only the momentum direction.

\subsection{Pressure isotropization}

\begin{figure}[t]
\begin{center}
\epsfig{file=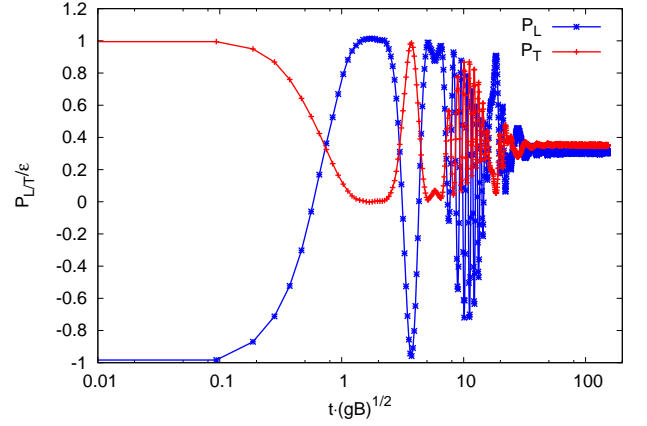, width=\mywidth, angle=270}
\caption{Dynamics of pressure isotropization for Nielsen-Olesen-type initial conditions.}
\label{fig:flux_isotrop}
\end{center}
\end{figure}

To answer the question of whether the observed decay of the coherent fields corresponds to an isotropization of the system, we investigate the time evolution of the diagonal elements of the symmetrized energy-momentum tensor $T^{\mu\nu}_x$, corresponding to the transverse and longitudinal pressure components
\begin{subequations}
\begin{align}
 \langle P_T \rangle&= -\frac{1}{2}\left\langle T^{1}_{\hspace{0.03cm}1,x}+T^{2}_{\hspace{0.03cm}2,x}\right\rangle\ , \\
 \langle P_L \rangle&= -\left\langle T^{3}_{\hspace{0.03cm}3,x}\right\rangle\ .
\end{align}
\end{subequations}
Isotropy would imply $\langle P_L\rangle=\langle P_T\rangle$, as is the case in thermal equilibrium. On the lattice, the pressure components take the form
\begin{subequations}
\begin{align}
 \langle P_T \rangle&= \frac{1}{2}\sum_{a}\left[\left(E^{a}_{3,x}\right)^2+\left(B^{a}_{3,x}\right)^2\right] \nonumber \\
                    &+\sum_{i=1}^{2}\frac{\operatorname{Im}\tr{\big[F_{x+\hat{i},x}(\gamma^i+\gamma_5)U_{i,x}-F_{x-\hat{i},x}(\gamma^i-\gamma_5){U^\dagger}_{\hspace{-0.15cm}i,x}\big]}}{2a_s} \, , \\
 \langle P_L \rangle&=\frac{1}{2}\sum_{a}\left[\sum_{i=1}^{2}\left[\left(E^{a}_{i,x}\right)^2+\left(B^{a}_{i,x}\right)^2\right]-\left(E^{a}_{3,x}\right)^2-\left(B^{a}_{3,x}\right)^2\right] \nonumber \\
		     &+\frac{\operatorname{Im}\tr{\big[F_{x+\hat{3},x}(\gamma^3+\gamma_5)U_{3,x}-F_{x-\hat{3},x}(\gamma^3-\gamma_5){U^\dagger}_{\hspace{-0.15cm}3,x}\big]}}{2a_s} \, .
\end{align}
\end{subequations}
In Fig.~\ref{fig:flux_isotrop} we show the time evolution of the volume-averaged longitudinal and transversal pressure components for Nielsen-Olesen-type initial conditions.
At early time, the different components show strong oscillations.
We emphasize, however, that the isotropic fixed point $P_L=P_T$ is reached at a time scale which corresponds to the decay time of the macroscopic field.

\begin{figure}[t]
\begin{center}
\epsfig{file=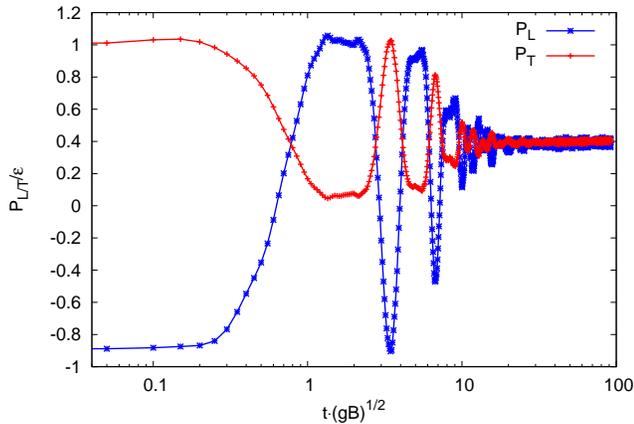, width=\mywidth, angle=270}
\caption{Dynamics of pressure isotropization for flux tube initial conditions.}
\label{fig:flux_isotrop2}
\end{center}
\end{figure}

In fact, the same behavior is found for flux tube initial conditions as shown in Fig.~\ref{fig:flux_isotrop2}. 
Again, both pressure components exhibit damped oscillations until they finally converge to the isotropic limit $P_L=P_T$.  
We note that the isotropization for flux tube initial conditions proceeds slightly faster than for condensate initial conditions.
This can be traced back to the rapid decay of the longitudinal chromoelectric field by producing quarks and gluons.

\subsection{Gluon distribution}

We have seen that both condensate and flux tube initial conditions result in pressure isotropization on time scales which characterize the decay times of the initial macroscopic fields. This means that all initial condition scenarios under consideration (overpopulation, Nielsen-Olesen-type, flux tube) lead to isotropy. In the following, we investigate whether the particle spectra and total particle numbers resulting from initial conditions with either a large coherent field or a large characteristic occupancy show similar behavior.

\begin{figure}[t]
\begin{center}
\epsfig{file=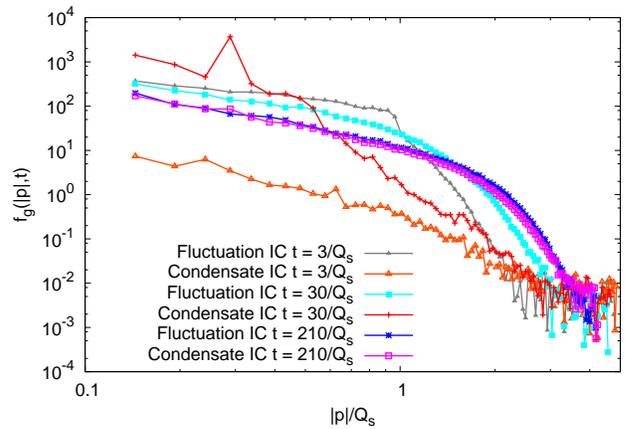, width=\mywidth, angle=270}
\caption{Gluon distribution $f_g(|\mathbf{p}|,t)$ at three different instants of time for gluonic overpopulation (``Fluctuation IC'') and Nielsen-Olesen-type initial conditions (``Condensate IC'').
The distributions emerging from the different initial conditions converge to the same isotropic form.}
\label{fig:gluon_spec1}
\end{center}
\end{figure}

To this end, we consider the particle distributions and particle numbers emerging from the Nielsen-Olesen instability and the gluonic overpopulation scenario.
To allow for a proper comparison, we choose comparable energy densities at initial times and employ identical numerical parameters (gauge coupling, lattice spacing, volume).
In Fig.~\ref{fig:gluon_spec1}, we present the gluon distribution for the two different scenarios at three different instants of time.

At early times $t = 3/Q_s$, the two distributions are clearly distinct:
For condensate initial conditions we find that the Nielsen-Olesen instability populates gluon modes in the infrared at a considerably higher rate than in the ultraviolet.
Nonetheless, most of the energy is still contained in the chromomagnetic field $B(t)$ such that the overall gluon occupancy is still rather small.
This is in contrast to the overpopulation scenario with an initial gluon occupation of the order of $1/g^2$ up to $|\mathbf{p}|=Q_s$.
After a short time, the rapid drop in occupancy around $|\mathbf{p}|=Q_s$ is still visible, however, gluon modes at somewhat higher momenta become populated as well.

At intermediate times $t = 30/Q_s$, the Nielsen-Olesen instability has fully developed by increasing the gluonic occupation in the infrared exponentially fast.
Most notably, this results in a gluon distribution at low momenta which becomes even higher than the initial occupation $1/g^2$ in the overpopulation scenario.
In fact, the gluon distribution in the overpopulation scenario has decreased for modes $|{\bf p}|\lesssim Q_s$. 

For later times around $t = 210/Q_s$ the two distributions have become almost indistinguishable and exhibit a power-law behavior towards the infrared with an approximate exponent $\kappa=3/2$.
At comparable time scales, this particular value of $\kappa$ has been found previously \cite{Berges:2012ev}.
At even later times, the power-law exponent is supposed to further decrease and approach $\kappa=4/3$ \cite{Berges:2008mr,Schlichting:2012es} before it becomes indistinguishable from a classical, thermal exponent $\kappa_{\text{th}}=1$ \cite{York:2014wja}.

\begin{figure}[b]
\begin{center}
\epsfig{file=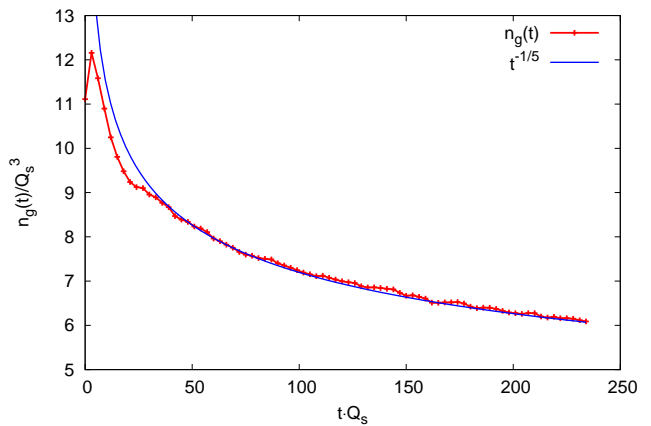, width=\mywidth, angle=270}
\caption{Self-similar time evolution of the total gluon number $n_g(t)$ for gluonic overpopulation initial conditions at $g^2N_f = 0.02$.}
\label{fig:totnum_powlaw_gluons}
\end{center}
\end{figure}

To emphasize the connection of our results to the nonthermal fixed point behavior found in \cite{Berges:2012ev,Berges:2014bba}, we consider the self-similarity relation
\begin{equation}
 f_g\left(|{\bf p}|,t\right) = t^{\alpha}f_S\big(t^{\beta}|{\bf p}|\big) \, ,
\end{equation}
which is valid for the gluon distribution in the scaling regime.
Here, $\alpha$ and $\beta$ are universal scaling exponents and $f_S$ is a time-independent scaling function.
We determine the scaling exponent $\beta$ by fitting the total number of gluons
\begin{equation}
 n_g(t)=\int\!\frac{d^3p}{(2\pi)^3}\,f_g(|{\bf p}|,t)
\end{equation}
with a power-law ansatz. 
In fact, the temporal scaling $n_g(t)\sim t^{\alpha-3\beta}n_g(0)$ in combination with energy conservation constrains the second scaling exponent, $\alpha=4\beta$ \cite{Berges:2014bba}.
As shown in Fig.~\ref{fig:totnum_powlaw_gluons}, we find for this scaling regime the exponent $\beta=-1/5$ to very good accuracy. 

\subsection{Quark distribution}

We have seen that the gluon distribution for both Nielsen-Olesen and gluonic overpopulation initial conditions become very similar at late times. The small differences between the two curves may be traced back to deviations in the total energy density and to the effects of quarks. In the following, we analyze the behavior in the quark sector.

\begin{figure}[t]
\begin{center}
\epsfig{file=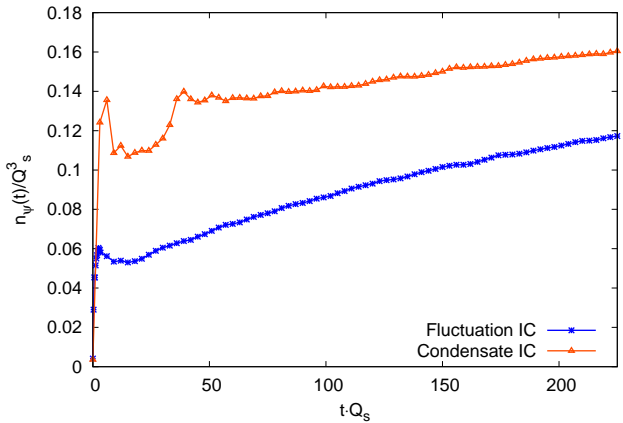, width=\mywidth, angle=270}
\caption{Time evolution of the total quark number density $n_\psi(t)$ for gluonic overpopulation and Nielsen-Olesen initial conditions.}
\label{fig:quark_totpart}
\end{center}
\end{figure}

In Fig.~\ref{fig:quark_totpart}, we show the time evolution of the total quark number
\begin{equation}
 n_{\psi}(t)=\int\!\frac{d^3p}{(2\pi)^3}\,f_{\psi}({\bf p},t)
\end{equation}
for both the Nielsen-Olesen-type instability and the gluonic overpopulation scenario.
One observes that the quark number increases abruptly at very early times in both cases.
The initial steep increase is caused by the free fermion vacuum initial condition together with the sudden switching-on of the coherent gauge fields for the Nielsen-Olesen-type initial conditions and of the high gluon occupancies in the overpopulation scenario, respectively. For given energy density, one observes that initially the quark production from coherent field decay is more efficient than quark production from gluon scattering. The rapidness of the initial production is also due to the very small quark masses $m\leq 10^{-2}Q_s$ and the effective absence of Pauli suppression for the low fermion occupancies at sufficiently early times.

\begin{figure}[t]
\begin{center}
\epsfig{file=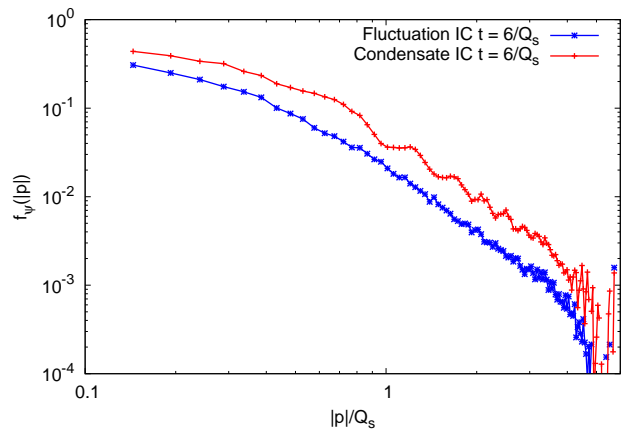, width=\mywidth, angle=270}
\caption{Quark distribution $f_\psi(\mathbf{p},t)$ at early times starting from gluon overpopulation and from Nielsen-Olesen-type initial conditions.}
\label{fig:quark_early_spec}
\end{center}
\end{figure}

After the characteristic decay time of the macroscopic field for Nielsen-Olesen-type initial conditions, the evolution of $n_\psi(t)$ exhibits a continuing but slower fermion production at later times for both scenarios. We observe from Fig.~\ref{fig:quark_early_spec} that the production rates become rather insensitive to the details of the initial state, which is in line with the universality in the gluon sector at those times. The quark production rates will be discussed further in section~\ref{subsec:qcd_largeNf}.   

\begin{figure}[t]
\begin{center}
 \epsfig{file=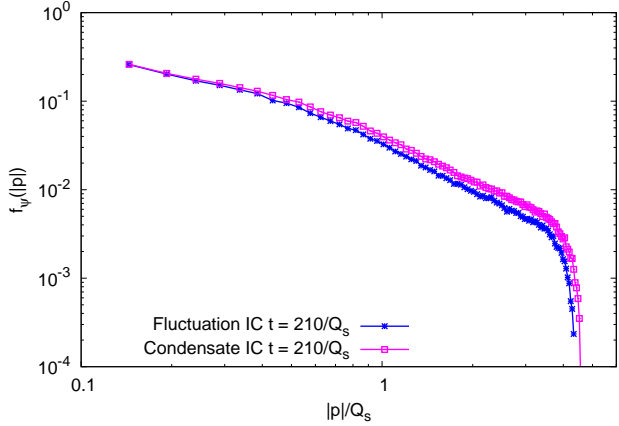, width=\mywidth, angle=270}
 \caption{Quark distribution $f_\psi(\mathbf{p},t)$ at later time $t = 210/Q_s$ starting from gluon overpopulation and from Nielsen-Olesen-type initial conditions. One observes the emergence of an approximate power-law behavior in the intermediate momentum regime. The values of the power-law exponents for both types of initial conditions are in the range $1.70 - 1.75$.}
\label{fig:quark_spec1}
\end{center}
\end{figure}

We have seen that the Nielsen-Olesen-type initial condition results in a larger value of the total quark number $n_\psi(t)$ than the overpopulation initial condition. In fact, this is reflected in the corresponding momentum-dependent quark distribution $f_\psi(\mathbf{p},t)$ at early times $t= 6/Q_s$ as shown in Fig.~\ref{fig:quark_early_spec}:
The spectrum resulting from initial overpopulation is lower in the whole momentum regime. In Fig.~\ref{fig:quark_spec1}, we compare the quark distributions at later time $t = 210/Q_s$. Even though there are still more quarks present starting from Nielsen-Olesen-type initial conditions, there is an apparent resemblance. Most notably, a power-law behavior in the intermediate momentum regime is found in both scenarios, suggesting the the late-time behavior becomes very similar. In fact, the value of the fermion power-law exponent is similar to the gluon exponent on similar time scales. A related phenomenon, where the quarks inherit approximate scaling properties of nonequilibrium bosons in an intermediate momentum regime, has been observed in a Yukawa theory before~\cite{Berges:2013oba}.

\begin{figure}[b]
\begin{center}
 \epsfig{file=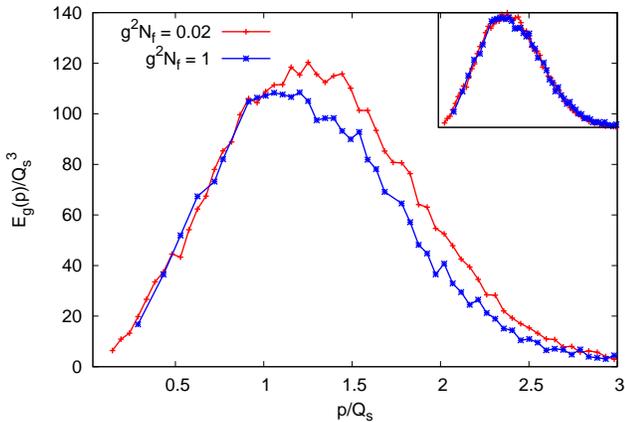, width=\mywidth, angle=270}
 \caption{Gluon quasi-particle energy distribution $E(|\mathbf{p}|)$ at $t = 60/Q_s$ for gluon overpopulation initial conditions for $g^2 N_f=1$ compared to $g^2 N_f=0.02$. In the inset we show that the rescaling of both width and height of the $g^2N_f = 0.02$ curve with a factor of $0.94$ causes both curves to practically overlap. In this and following plots the results for $g^2N_f = 1$ have been obtained from averaging over four runs.}
\label{fig:energ_gluon_spectra}
\end{center}
\end{figure}

\section{Quark backreaction and large $N_f$}
\label{subsec:qcd_largeNf}
In the previous section we studied $N_f=2$ degenerate light quark flavors in the weak coupling regime with $g^{2}=10^{-2}$. While the initial large gluon fields or occupancies are found to have a dramatic impact on quark production, the backreaction of the quark sector on the gluon distribution has only the expected minor consequences at weak coupling. However, what controls this backreaction is the product of the coupling squared and the number of quark flavors such that for fixed $g^2N_f$ of order one even the weak-coupling limit becomes strongly correlated for a large enough number of flavors. This opens the striking possibility to simulate strong-interaction aspects while staying within the range of validity of our real-time lattice simulation techniques.  

We emphasize, however, that the corresponding numerical simulations require substantially larger resources as the number $N_{\text{sto}}$ of pairs of male and female spinor fields has to be significantly increased in this case. As a consequence, we perform our numerical simulations on a somewhat smaller $32^3$ spatial lattice for $g^2N_f=1$ as compared to the $64^3$ spatial lattice for $g^2N_f=2\cdot10^{-2}$.
In the following, we restrict ourselves to either gluon overpopulation or Nielsen-Olesen-type initial conditions.

\begin{figure}[t]
\begin{center}
 \epsfig{file=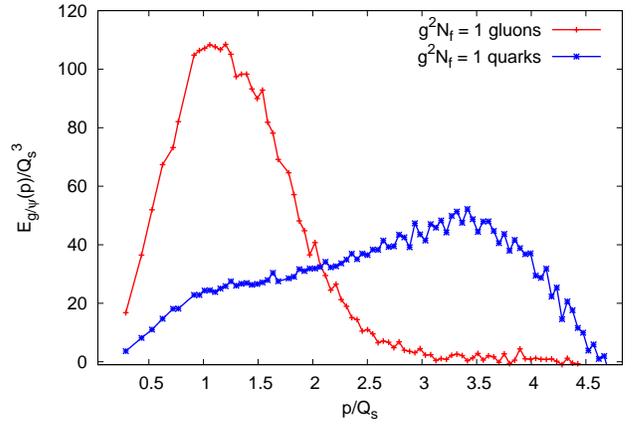, width=\mywidth, angle=270}
 \caption{Gluon and quark quasi-particle energy distributions at $t = 60/Q_s$ for $g^2 N_f=1$ with gluon overpopulation initial conditions.}
\label{fig:energ_quark_gluon_spectra}
\end{center}
\end{figure}

In Fig.~\ref{fig:energ_gluon_spectra} we study how the quark backreaction influences the gluon quasi-particle energy distribution
\begin{equation}
 E_g(|\mathbf{p}|)=6 p^3 n_g(|\mathbf{p}|) \, 
\end{equation}
for different numbers of quark flavors and fixed $g^{2}=10^{-2}$.
One observes that the peak gluon distribution is reduced for $g^2 N_f=1$ as compared to $g^2 N_f=0.02$. However, most remarkably we find that the shape of the gluon distribution is almost unaltered. More precisely, the rescaling of both width and height by a common factor causes both curves to practically overlap as demonstrated in the inset of Fig.~\ref{fig:energ_gluon_spectra}. As a consequence, the universal properties of the gluon sector endure the impact of strong quark backreactions, at least at the level of accuracy considered.

In Fig.~\ref{fig:energ_quark_gluon_spectra}, we show the quasi-particle energy distribution of quarks
\begin{equation}
 E_{\psi}(|\mathbf{p}|)=8N_f p^3 f_{\psi}(|\mathbf{p}|) \, .
\end{equation}
In contrast to the gluon distribution, whose peak is around $p \simeq Q_s$, the dominant quark momenta have shifted to higher values. In contrast to gluons, which can be highly occupied, the quark occupancies are limited by the exclusion principle such that more and more states at higher momenta have to be filled to account for the increased energy in the quark sector as $g^2 N_f$ is enlarged.  
These findings resemble results for Yukawa theories, where a similar separation of momenta was found~\cite{Berges:2013oba}. Apparently, increasing $g^2 N_f$ results in an efficient mechanism of transporting energy from the infrared to the ultraviolet, which should lead to a quickening of kinetic equilibration.

\begin{figure}[t]
\begin{center}
 \epsfig{file=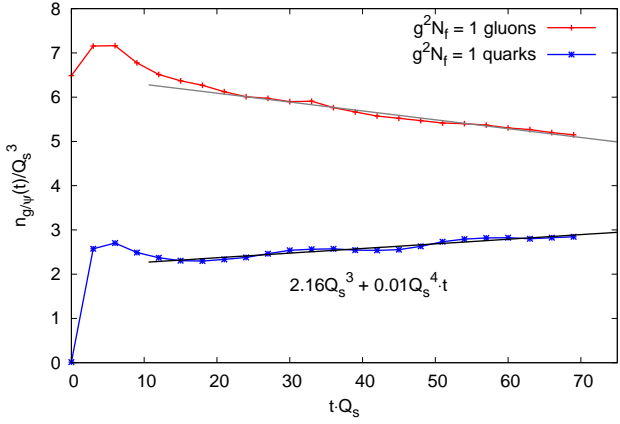, width=\mywidth, angle=270}
 \caption{Time evolution of total gluon number $n_g(t)$ and total quark number $n_\psi(t)$ for $g^2N_f=1$.}
\label{fig:totnum_quark_fit_back}
\end{center}
\end{figure}

In order to achieve chemical equilibration, the total number of produced quarks $n_\psi(t)$ needs to reach its thermal equilibrium value. 
In view of the limitations of the current methods to describe the long-time behavior, we consider here a rough estimate of the chemical equilibration time based on an extrapolation of the quark production rate.

To this end, we perform a linear fit of the time-dependent particle number
\begin{subequations}
\begin{align}
 n_\psi(t)^{g^2 N_f=0.02}&=0.055Q_s^3+0.0003Q_s^4 t \ , \\
 n_\psi(t)^{g^2 N_f=1}&=2.16Q_s^3+0.01Q_s^4 t \ ,
\end{align}
\end{subequations}
the latter being presented in Fig.~\ref{fig:totnum_quark_fit_back}.
To estimate the chemical equilibration time, we seek the point at which agreement with the Stefan-Boltzmann limit of massless non-interacting quasi-particles
\begin{equation}
n^{SB}_{\psi} = \frac{6\zeta(3)}{\pi^2}N_f T^3_{SB} \, ,
\end{equation}
with the Riemann zeta-function $\zeta(x)$, is obtained.
In fact, the final temperature $T_{SB}$ of the closed system can be determined from energy conservation.
For gluon overpopulation initial conditions, for instance, equating the energy density in the Stefan-Boltzmann limit with the energy density of the initial state
\begin{equation}
 \frac{\pi^2}{30}(6+7N_f)T_{SB}^4 \stackrel{!}{=}\frac{3}{4\pi^2}\frac{Q_s^4}{g^2} \ ,
\end{equation}
results in 
\begin{equation}
T_{SB} = \frac{Q_s}{\pi}\left(\frac{45}{2(6+7N_f)g^2}\right)^{1/4} \, .
\end{equation}

The corresponding estimate for the chemical equilibration time $t_{ch}$ then gives $t_{ch}^{g^2 N_f=0.02}\simeq 4500/Q_s$ and $t_{ch}^{g^2 N_f=1}\simeq 345/Q_s$, respectively. Without taking into account the backreaction of quarks onto gluons, one would expect the total quark production rates to be proportional to the number of degenerate flavors for fixed coupling. However, we find
\begin{equation}
 13\simeq\frac{t_{ch}^{(g^2 N_f=0.02)}}{t_{ch}^{(g^2 N_f=1)}} < \frac{N_f = {1/g^2}}{N_f = 2}=50 \ ,
\end{equation}
which points to the reduction of scattering rates by the diminished gluon occupation numbers for enhanced quark backreactions.

\section{Conclusions}
\label{sec:qcd_conclusions}

We have studied gluon dynamics and quark production in two-color QCD with light quarks from simulations in $3+1$ dimensional space-time. We concentrated on the weak-coupling regime in order to guarantee the applicability of 
the classical-statistical approximation for the gluon sector, while we simulated the quark dynamics in a stochastic approach taking into account the quantum nature of the fermions. In order to understand the importance of fermion backreaction, we studied the dynamics for different $g^2 N_f$ by changing the number of quark flavors for fixed coupling. 

Having considered three types of initial conditions with large fields or occupancies for $N_f = 2$, we confirmed the universality of the dynamics near the non-thermal fixed point existing in gluon systems. The scaling exponents we found in the gluon sector are in agreement with earlier investigations in pure Yang-Mills simulations. We demonstrated that anisotropic initial conditions leading to plasma instabilities and Schwinger pair production in QCD isotropize rather quickly for the non-expanding system, such that their subsequent approach to thermal equilibrium via energy cascade to short length scales can be described in an entirely isotropic framework.

The universality of the gluon dynamics has profound effects on quark production. Although very different at early stages, the total numbers and spectral distributions of quarks produced from the different gluon initial conditions considered
tend to approach each other at later times. The corresponding spectral distributions of quark particle numbers acquire a shape marked by a power-law at intermediate momenta in the weak-coupling regime for small $g^2 N_f$. 

While the initial large gluon fields or occupancies are found to have a dramatic impact on quark production, the backreaction of the quark sector on the gluon distribution has only minor consequences for small $g^2 N_f$. We increased the number of flavors such that $g^2 N_f$ becomes of order one leading to a strongly correlated fermion sector. Most remarkably, we find that the shape of the gluon distribution is almost unaltered by changing $N_f$ despite the fact that the peak gluon distribution can be reduced considerably. Furthermore, for a strongly correlated fermion sector the dominant quark momenta have shifted to significantly larger values as compared to the characteristic gluon momentum of $p \simeq Q_s$. Our estimates for the chemical equilibration time indicate that calculations neglecting the backreaction for strongly correlated quarks lead to substantially shorter times than a full calculation taking into account the impact on the gluon sector. Interestingly, a prolonged chemical equilibration could have a phenomenological significance by contributing to the elliptic flow of thermal photons~\cite{Monnai:2014kqa}, a quantity which appears to be underestimated by hydrodynamic models~\cite{Adare:2011zr,Lohner:2012ct}.

In view of applications to heavy-ion collisions, taking $g^2 N_f$ of order one is expected to be a reasonable assumption and our results provide important insights into nonequilibrium QCD dynamics from first principles. The next step would be to include the longitudinal expansion of the plasma's space-time evolution.\\


We thank R.~Alkofer, V.~Kasper, N.~M\"uller, S.~Schlichting, N.~Tanji and R.~Venugopalan for discussions. 
This work is supported by the DFG and  
D.~Gelfand thanks Austrian Science Fund (FWF): P 26582-N27 and HGS-HIRe for FAIR for their support.
F.~Hebenstreit acknowledges support from the Alexander von Humboldt Foundation in the early stages of this work as well as from the European Research Council under the European Union's Seventh Framework Programme (FP7/2007-2013)/ ERC grant agreement 339220.

\end{document}